# Energy flow structuring in the focused field

Hao Chen[1] and Guoqiang Li[1,2,*]

[1]Department of Ophthalmology and Visual Science, The Ohio State Universit,y, Columbus, OH 43212, USA
[2]Department of Electrical and Computer Engineering, The Ohio State University, Columbus, OH 43212, USA

[*]Corresponding author: li.3090@osu.edu

Abstract

We propose an iterative method of energy flow shaping in the focal region with the amplitude, phase and polarization modulation of incident light. By using an iterative optimization based on the diffraction calculation with help of the fast Fourier transform, we can tailor the polarization and phase structure in the focal plane. By appropriate design of the polarization and phase gradients, arbitrary energy flow including spin and orbital parts can be designed and tailored independently. The capability of energy flow structuring is demonstrated by the measurement of the Stokes parameters and self-interference pattern. This provides a novel method to control the vectorial feature of the focal volume.

Poynting vector can be divided into a spin part and an orbital part associated with the polarization and phase spatial inhomogeneity of an optical field [1]. These two sorts of energy flow can be transferred to atoms, molecules, and particles, resulting in optical force that pushes objects along the flow direction [2-7]. Optical force stemming from phase gradient has been widely discussed and used in optical tweezers, such as optical spanners, optical pumps that create directed fluid flows [4,5]. The other part of nonconservative scattering force energy flow associated with the curl of polarization is observed recently in the diffusion of a metal nanoparticle and can be used for trapping and guiding nanoparticles [8, 9]. However, generation and manipulation of both parts of energy flow simultaneously in the focused field still remains a challenging task. Moreover, there has not been much effort devoted to simultaneous control of polarization state and phase structure of the focal field [10]. In this Letter, we propose an iterative method for energy flow shaping in the focal region through proper design of polarization and phase gradient and search for the optimized amplitude, phase and polarization modulation of the incident field. This method is based on the scalar diffraction with low numerical aperture (NA) and can be easily extended to the situation with high NA microscope system by introducing vectorial diffraction. It also has potential applications in optical tweezers by manipulating both spin force and orbital force [11-13].

The vector potential for a monochromatic beam of light with angular frequency $\omega$ is written as $\vec{\mathbf{A}}(\mathbf{r},t)=u(\mathbf{r})e^{i\varphi(\mathbf{r})-i\omega t}\hat{\boldsymbol{\varepsilon}}(\mathbf{r})$ [2], where $u(\mathbf{r})$ is the real-valued amplitude, $\varphi(\mathbf{r})$ is the real-valued phase, and $\hat{\boldsymbol{\varepsilon}}(\mathbf{r})$ is the complex-valued polarization vector at position *r*. According to the Poynting's theorem, the time-averaged Poynting vector also expresses the momentum density written as [1, 14]

$$\vec{\mathbf{p}} = \frac{1}{c^2}\vec{\mathbf{S}} = \frac{g}{2c^2}\mathrm{Re}[\vec{\mathbf{E}}^* \times \vec{\mathbf{H}}], \qquad (1)$$

where *g* is a constant factor and equals to $(8\pi)^{-1}$, and *c* is the velocity of light. From now on, the pre-factor $\frac{g}{2c^2}$ will be ignored. By using the Maxwell equations, the momentum density can be rewritten as

$$\vec{p} = \text{Im}\left[\vec{E}^* \times (\nabla \times \vec{E})\right] = \vec{p}_o + \vec{p}_s, \tag{2}$$

where
$$\vec{p}_o = \text{Im}(\vec{E}^* \cdot \nabla \vec{E}) = I \cdot \nabla \varphi, \tag{3}$$

$$\vec{p}_s = \nabla \times \vec{s}. \tag{4}$$

And $\vec{s} = i \cdot \vec{E} \times \vec{E}^* = i(\vec{E}_x \vec{E}_y^* - \vec{E}_y \vec{E}_x^*) = I_+ - I_-$ (5)

is the spin angular momentum and is also characterized as the degree of circular polarization. Correspondingly, $z$ coincides with the propagation direction, $x$ and $y$ are the transverse coordinates. The momentum density is decomposed into two terms, which refer to orbital flow density (OFD) and spin flow density (SFD). The OFD is associated with phase gradient of the optical field and the SFD is associated with polarization gradient [15,16]. The difference lies in that the direction of OFD is identical to the phase gradient while the direction of SFD is orthogonal to the polarization gradient. How to realize the arbitrary designed $\vec{p}_o$ and $\vec{p}_s$ in the focused field by the incident field is the main target of this paper.

Our endeavor starts with an inhomogeneous beam with randomly distributed amplitude, phase and polarization in the cross section of beam, described as $A(x,y)e^{i\phi(x,y)}(\cos\varphi(x,y), \sin\varphi(x,y))^T$, where $A(x,y)$, $\phi(x,y)$ and $\varphi(x,y)$ are the amplitude, phase retardation and polarization angle of the incident light. In this way, three parameters of the incident light need to be optimized during the iterative procedure for generating a desired polarization gradient and phase gradient structure in focus [17,18]. Accordingly, the corresponding spin flow and orbital flow are produced in the focal region with arbitrary structure.

Firstly, the incident field $E_{in}(x, y)$ is divided into two components in the helicity basis $(\hat{e}_+ = [1,i]^T, \hat{e}_- = [1,-i]^T)$ for simplicity. We have

$$E_{in}(x,y) = A(x,y)e^{i\phi(x,y)}\left[\cos\varphi(x,y), \sin\varphi(x,y)\right]^T = A(x,y)[1,i]^T e^{i\phi_1(x,y)} + A(x,y)[1,-i]^T e^{i\phi_2(x,y)}, \tag{6}$$

where $\phi_1(x,y) = \phi + \varphi$ and $\phi_2(x,y) = \phi - \varphi$. The focal field of a monochromatic light passing through an aplanatic lens can be calculated by the Fresnel diffraction integral and rewritten in terms of the Fourier transform [19]:

$$E_f(u,v) = \begin{bmatrix} E_+(u,v) \\ E_-(u,v) \end{bmatrix} = \int\int_{-\infty}^{\infty} \begin{bmatrix} A(x,y)e^{i\phi_1(x,y)} \\ A(x,y)e^{i\phi_2(x,y)} \end{bmatrix} \exp\left[-j2\pi\left(\frac{x}{\lambda f}x_0 + \frac{y}{\lambda f}y_0\right)\right] dx_0 dy_0. \quad (7)$$

We can make a fast calculation for the helicity components of the field in the focal region by using the FFT operations. In order to obtain controlled formation of energy flow distribution, we restrict our attention to how to realize a special distribution of polarization and phase in the focal plane by modifying the three parameters of the incident beam. Our scheme simulates an iterative process between the forward and backward propagations that relate the incident and the focused fields. This process is analogous to the Gerchberg-Saxton algorithm for retrieving the phase of a pair of light distributions related via the Fourier transform [20]. The flow chart of the iterative algorithm is depicted in Fig. 1. The iterative process starts with assuming a set of randomly distributed amplitude and phase of two components of the incident beam (i.e., $\phi_1$ and $\phi_2$ in Eq. (7) are two irrelevant uniform random number in [0, $2\pi$], $A$ is uniform random number in [0, 1]). The diffraction integrals [Eq. (7)] are then calculated using the FFT, yielding the helicity components of the field in the focal plane. Both of the absolute values and phase values of the calculated helicity components in the focal field are replaced by the prescribed distribution. Two new distributions in the incident beam are obtained through an inverse FFT by letting the reshaped focal field propagate backward to the entrance plane. A constraint that both of the helicity components in the incident field have uniform amplitude is imposed on the renewed incident beam, indicated by Eq. (7). Hence, the renewed amplitude distribution is set by $A_i(x,y) = \left[A'_{i+}(x,y) + A'_{i-}(x,y)\right]/2$, while the phase values are left unchanged. The iterative search continues until a satisfactory focused field is found. It should be mentioned that longitudinal components of the optical field is omitted due to the low NA situation.

In a previous paper [21], an interferometric method has been developed to generate vector beams with arbitrary distribution of phase and polarization. A 632.8nm linear polarization light is input into the spatial light modulator (SLM)-based 4f setup with spatial filtering, giving rise to an output of the vector beams. A computer controls the specially designed pattern that is projected onto the SLM, enabling the dynamic generation of a space-variant linear polarization state and phase structure (refer to Ref. [21] for details),

Furthermore, additional amplitude modulation can be addressed with controlled the modulation depth function [10]. Now we utilize this optical system to create the vector beams that are optimally designed by the above-described iterative search. The generated vector beams are then focused by a lens with a focal length of 150 mm.

Here we present examples to demonstrate vectorial control of the focused field. The first two examples are shown in Figs. 2 and 3. Both of the examples are double-mode focused fields. The inner mode of the first example (Fig. 2) is radially polarized and the outer mode is linearly polarized with helical phase structure. The inner mode of the second example (Fig. 3) is azimuthally variant and the outer mode is radially variant polarized structure. Figures 2(b) and 3(b) illustrate the polarization states in the focal plane. In both figures, the intensity of the left-, and right-circularly polarized components are shown in Figs. 2(c-d) and Figs. 3 (c-d), and the phase structure of the left-, right-circularly polarized components are shown in Figs. 2 (e-f) and Figs. 3 (e-f), respectively. Moreover, the Stokes parameters (Figs. 2 (g-j)) and Figs. 3 (g-j)) are demonstrated to verify the control of the focused vectorial field.

Also we present additional examples to demonstrate the directed energy flow in the focal plane by modulating the incident beam. As elucidated above, the spin flow is orthogonal to the polarization gradient and the orbital flow is parallel with the phase gradient. Arbitrarily directed energy flow can be obtained by specially designed polarization and phase structure. The first example is shown in Fig. 4. Figure 4(d) illustrates the polarization state in the focal plane, wherein the inner ring is x-polarized and the outer ring has polarization gradient in the radial direction. The simulated intensity and phase of left-, right-circularly polarized components are also shown in Figs. 4(a-b, e-f). According to the simulated optical field distribution, the spin and orbital flow density of the focused field is depicted in Figs. 4 (m) and (n), respectively. The blue arrows indicate the direction of each flow. The different types of flow density are spatially divided in the separated ring-shaped region. Oppositely directed spin flow is also shown on the edge of the outer ring. The phenomenon is due to the fact that the spin circulation cells do not compensate at the edge of the ring. Furthermore, measurements of the self-interference pattern (Figs. 4 (g-h)) and the Stokes parameters (Figs. 4 (i-l)) are explored to verify the polarization structure and the phase structure of the focused field. In order to get insights into the phase structure, we place an amplitude grating in front of the CCD camera so as to create a self-interference pattern on the CCD camera. As it is known that the dielectric grating is polarization-insensitive, the interference pattern is related to the phase structure. The generated vector field illuminates the amplitude grating which diffracts the incoming light into various diffraction orders. When the frequency of the grating is low, the interference pattern is generated by the slightly detached diffraction orders interfering with each other. The interference pattern with straight fringes corresponds to the vector beam with planar wavefront for comparison with the example below. In doing so, we do not directly observe the phase of the vector beam itself but rather indirectly confirm its existence through the self-interference fringes behind the amplitude grating resulting from superposition of various diffraction orders. The second example is a focused pair of lines. All the detailed results are shown in Fig. 5. Each panel is presented in the same order as in Fig. 4. The left line has only orbital flow directed to the negative direction of y axis, while the right line has only spin flow directed to the negative direction of y axis.

To summarize, we have proposed an approach using iterative method with the FFT to reshape the polarization component in the focal plane. This provides a novel way to tune the polarization states and phase structure in the focal volume with the help of the incident field modulation. Furthermore, the spin and orbital flows in the focused field could be manipulated independently and dynamically. The scheme we presented can be easily extended to the case with high NA focusing lens by a vectorial integral, and it is also very valuable to expand the functionality of energy flow in the optical tweezers.

**Figure captions**

Fig. 1. Flow chart for iteratively searching for the optimal amplitude, phase, and polarization modulation of incident beam.

Fig. 2. Double ring-shaped focused field. (a-b) Intensity of total field and the polarization state. (c-d) Intensity of the left-, and right-circularly polarized components; (e-f) Phase structure of the left-, right-circularly polarized components; (g-j) Stokes parameters $S_0$, $S_1$, $S_2$ and $S_3$ of the focused field.

Fig. 3. Double ring-shaped focused field. (a-b) Intensity of total field and the polarization state. (c-d) Intensity of the left-, and right-circularly polarized components; (e-f) Phase structure of the left-, right-circularly polarized components; (g-j) Stokes parameters $S_0$, $S_1$, $S_2$ and $S_3$ of the focused field.

Fig. 4. Double ring-shaped focused field. (a-c), Intensity of left-, right- circularly polarized components and the total field; (d) The polarization state of the focused field; (e-f) Phase structure of the left-, and right-circularly polarized components; (g-h) The self-interference pattern of the focused field — (g) Helical phase of the inner mode; (h) Planar phase of the inner mode; (i-l) Stokes parameters $S_0$, $S_1$, $S_2$ and $S_3$ of the focused field; (m-n) Spin and orbital flow density in the focused field.

Fig. 5. Double line-shaped focused field. (a-c), Intensity of left-, right- circularly polarized components and the total field; (d) The polarization state of the focused field; (e-f) Phase structure of the left-, and right-circularly polarized components; (g-h) The self-interference pattern of the focused field; (i-l) Stokes parameters $S_0$, $S_1$, $S_2$ and $S_3$ of the focused field; (m-n) Spin and orbital flow density in the focused field.

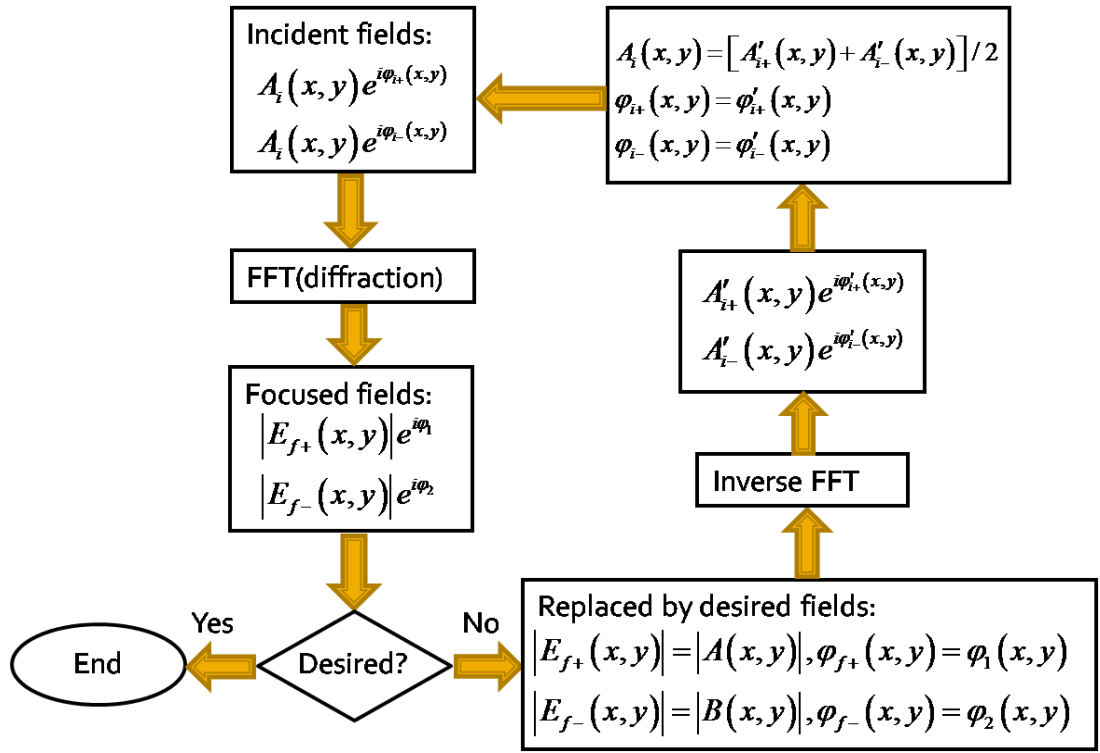

Fig. 1

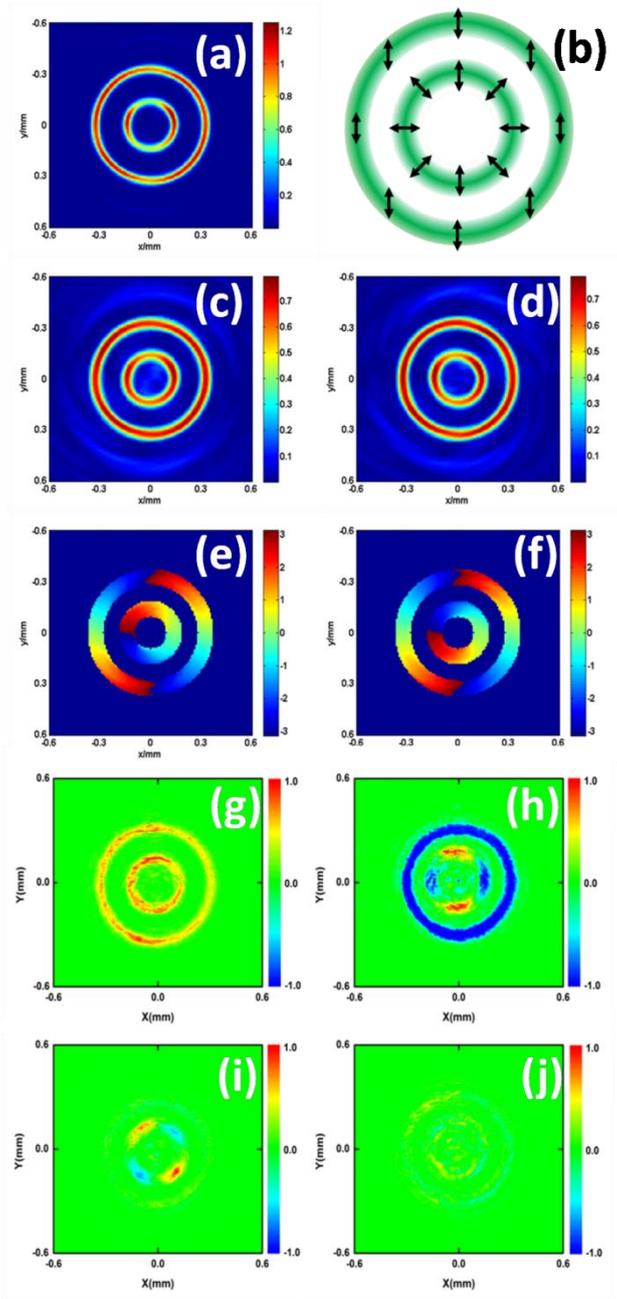

Fig. 2

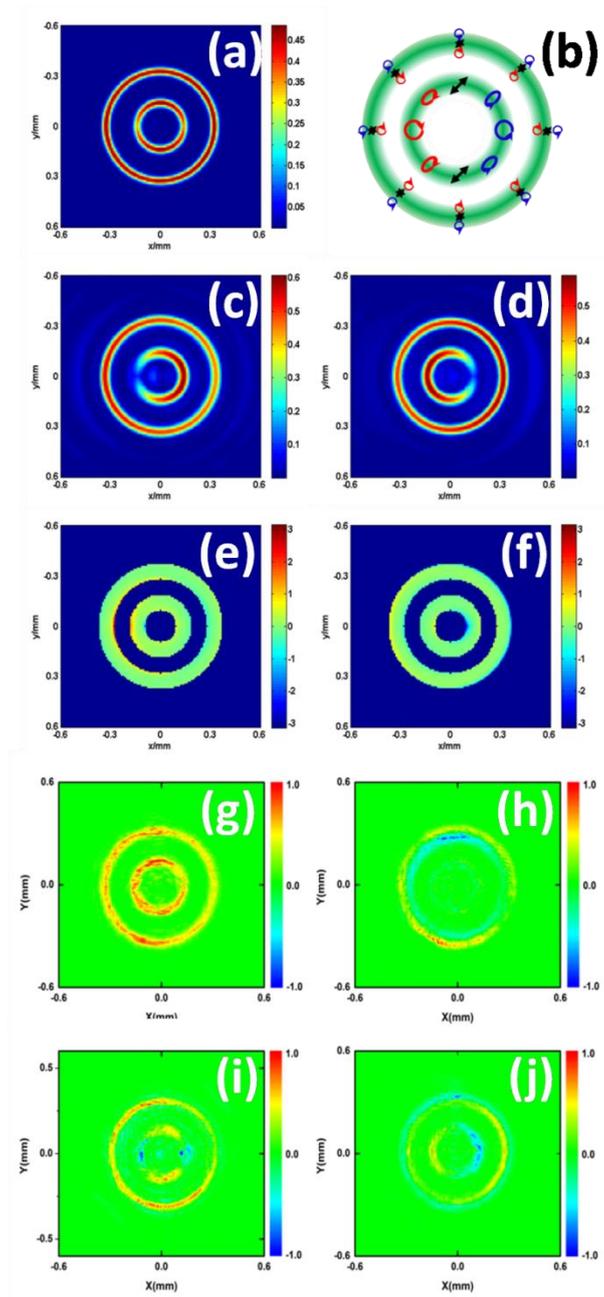

Fig. 3

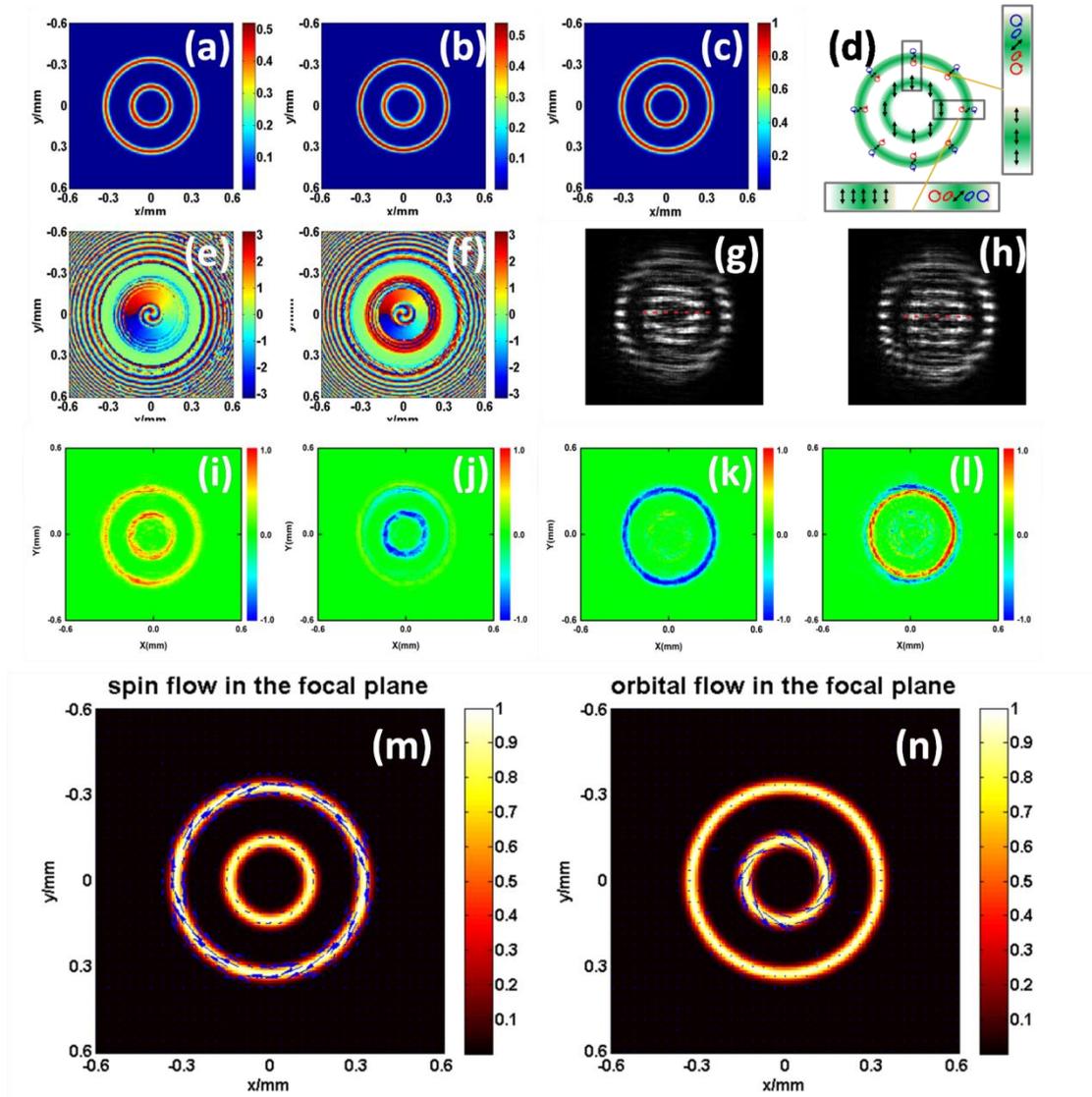

Fig. 4

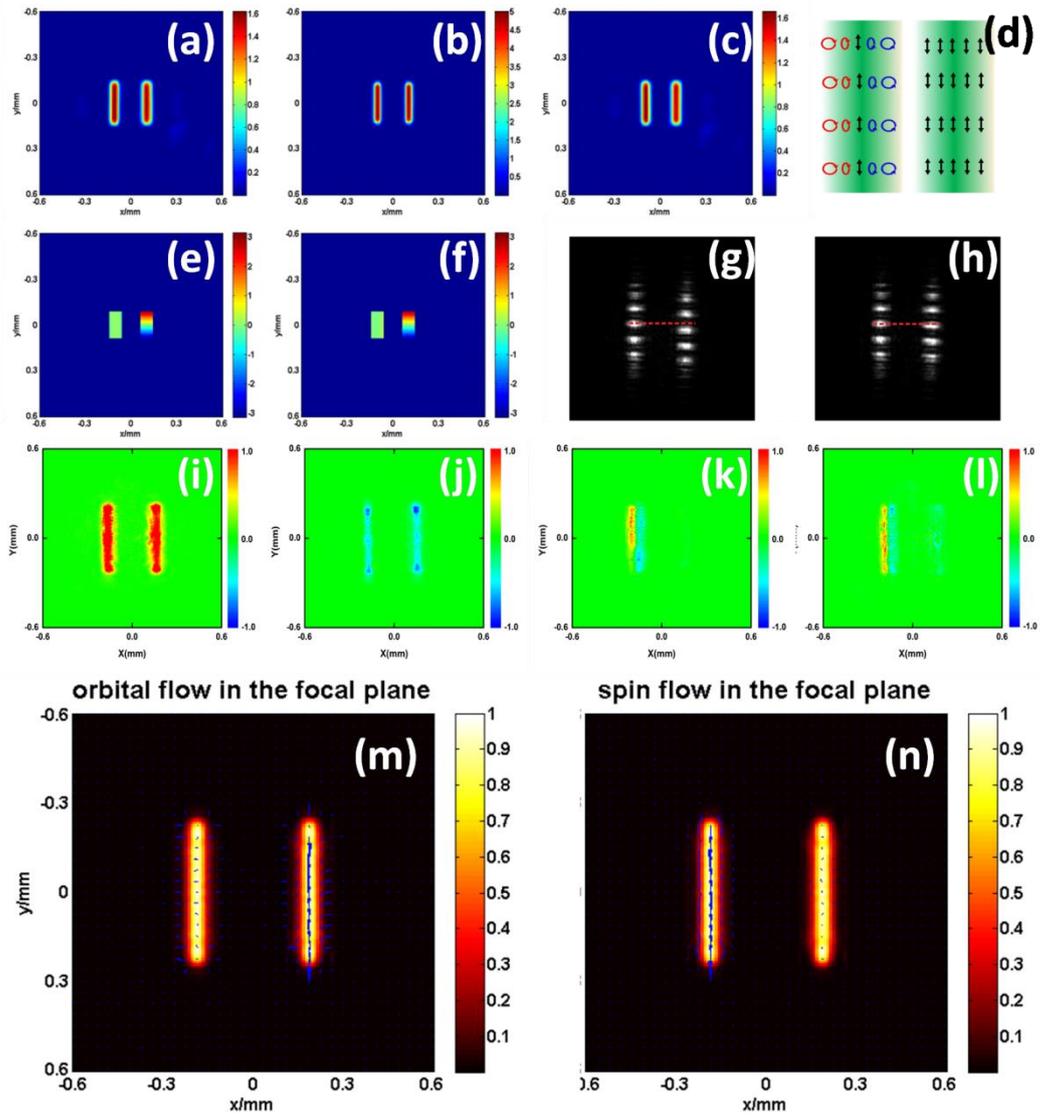

Fig. 5